\documentclass[reprint,amssymb,amsmath,aip,cha,twocolumn, superscriptaddress,frontmatterverbose]{revtex4-2}
\usepackage{graphicx}
\usepackage{dcolumn}
\usepackage{bm}
\usepackage{subfigure}
\usepackage{rotating}
\usepackage{epsfig}
\usepackage[normalem]{ulem}

\usepackage{lineno}
\usepackage{color}

\begin{document}


\title{Scale-free avalanches in arrays of FitzHugh-Nagumo oscillators}

\author{Max Contreras}
\email{mecontrl@gmx.de}
\affiliation{Institut f\"ur Theoretische Physik, Technische Universit\"at Berlin, Hardenbergstra\ss e 36, 10623 Berlin, Germany}

\author{Everton S. Medeiros}
\affiliation{Institute for Chemistry and Biology of the Marine Environment, Carl von Ossietzky University Oldenburg, 26111 Oldenburg, Germany}

\author{Anna Zakharova}
\affiliation{Institut f\"ur Theoretische Physik, Technische Universit\"at Berlin, Hardenbergstra\ss e 36, 
10623 Berlin, Germany}
\affiliation{Bernstein Center for Computational Neuroscience, Humboldt-Universit\"at zu Berlin, 
Philippstra\ss e 13, 10115 Berlin, Germany}

\author{Philipp H\"ovel}
\email{philipp.hoevel@uni-saarland.de}
\affiliation{Theoretical Physics and Center for Biophysics, Saarland University, Campus E2 6, 66123 Saarbr\"ucken, Germany}

\author{Igor Franovi\'c}
\email{franovic@ipb.ac.rs}
\affiliation{Scientific  Computing  Laboratory,  Center  for  the  Study  of  Complex  Systems,Institute  of  Physics  Belgrade,  University  of  Belgrade,  Pregrevica  118,  11080  Belgrade,  Serbia}

\date{\today}

\begin{abstract}
The activity in the brain cortex remarkably shows a simultaneous presence of robust collective oscillations and neuronal avalanches, where intermittent bursts of pseudo-synchronous spiking are interspersed with long periods of quiescence. The mechanisms allowing for such a coexistence are still a matter of an intensive debate. Here, we demonstrate that avalanche activity patterns can emerge in a rather simple model of an array of diffusively coupled neural oscillators with multiple timescale local dynamics in vicinity of a canard transition. The avalanches coexist with the fully synchronous state where the units perform relaxation oscillations. We show that the mechanism behind the avalanches is based on an inhibitory effect of interactions, which may quench the spiking of units due to an interplay with the maximal canard. The avalanche activity bears certain heralds of criticality, including scale-invariant distributions of event sizes.
Furthermore, the system shows an increased sensitivity to perturbations, manifested as critical slowing down and a reduced resilience.
\end{abstract}

\maketitle
\begin{quotation}
Cascading dynamics are a prominent feature of many complex systems, from information or disease spreading in social interactions to propagation of neuronal activity. Since the discovery of neuronal avalanches, it has been suggested that the brain cortex operates at criticality, leveraging this feature to maximize its dynamic range, information capacity, and dynamical repertoire. Nevertheless, in neuronal systems, the patterns of transient synchrony, such as avalanches, typically coexist and/or interact with robust collective rhythms, and the problem of generic mechanisms that give rise to avalanches and simultaneously allow for their coexistence with collective oscillations still remains unresolved. Here, we demonstrate that the avalanche activity can emerge and coexist with synchronous oscillations in a simple model of diffusively coupled neural oscillators with multiple timescale local dynamics in the vicinity of a canard transition. The avalanches are characterized by scale-invariant distributions of event sizes and an analysis of laminar, that is, inter-event, times. The latter quantifies both cascading and non-successive avalanches. At the critical transition between the states of lower and higher spiking rates that facilitates the onset of avalanches, the system exhibits an increased sensitivity to perturbations, manifested as critical slowing down and a reduced resilience. The disclosed scenario for coexistence of a well-defined oscillation rhythm and patterns with scale-invariant features may open a new avenue of research concerning multistability (and metastability) in neuronal systems.
\end{quotation}

\section{Introduction}\label{sec:intro}

The notions of criticality and phase transitions have gained a revived interest following the formulation of the concept of critical transitions and tipping \cite{Scheffer2009,Scheffer2012,Kuehn2011}, which essentially translate the ideas of bifurcation theory to the realm of complex systems. Naturally, the latter is not a straightforward process due to high-dimensional dynamics of complex systems. Moreover, in many applications, understanding the details of the states involved in a critical transition, as well as finding appropriate indicators of tipping proves as a difficult problem. Many complex systems exhibit multistability and metastability, an ample example being the brain activity. On the one hand, the functionality of the brain relies on generating robust collective rhythms based on synchronization at different levels of self-organization within the cortex \cite{Buzsaki2004,Buzsaki2006}. On the other hand, various types of experiments, both under in vivo and in vitro conditions, have revealed the presence of neuronal avalanches \cite{Beggs2003,Beggs2004,Plenz2007,Hahn2010}, that is, cascades of quasi-synchronous bursts of activity, whose main feature is scale invariance where the spatial and temporal distributions of events follow power-law behaviors. The discovery of neuronal avalanches has led to the brain criticality hypothesis \cite{Kinouchi2006,Chialvo2010,Shew2013,Cocchi2017,Munoz2018}, suggesting that the emergent cortical dynamics derive from being poised at the boundary of instability or at the edge of chaos. However, the precise character of the underlying continuous phase transition remains elusive \cite{Mariani2022,Fontenele2019,DallaPorta2019,diSanto2018}. Moreover, a question that naturally arises is how can so different types of activity, in particular those with a well-defined characteristic timescale (regular synchronous activity) and others where such timescales are absent (irregular transiently synchronous activity), coexist. Furthermore, what are the mechanisms that facilitate such a coexistence?

Recalling the classical theory of phase transitions, power-law behaviors should naturally be expected in scenarios where critical transitions can be associated with supercritical bifurcations. For instance, it is typically stated that neuronal avalanches emerge in the vicinity of a critical transition between silent (absorbing) and active states from a critical branching process \cite{Plenz2014, Chialvo2010} or at the synchronization transition \cite{Gireesh2008,Yang2012,diSanto2018,DallaPorta2019,Fontenele2019}. Nevertheless, power laws and other heralds of criticality, such as critical slowing down, have also been observed in relation to first-order phase transitions \cite{Levina2009,Millman2010}, where criticality involves multistable and metastable behavior. This also applies to certain models of neuronal avalanches, which have indicated their onset in the vicinity of a discontinuous transition showing hysteresis between the low-activity (down) and the high-activity (up) state \cite{Scarpetta2018}. Nevertheless, the general mechanisms that can reconcile the emergence of avalanche-like patterns with collective rhythms in neuronal systems, are still a subject of on-going research \cite{Miller2019,DallaPorta2019,diSanto2017,Plenz2014}.

Motivated by the latter problem, we show in this paper that avalanche-like bursting patterns can emerge in a rather simple model of an array of non-locally coupled FitzHugh-Nagumo (FHN) units with attractive diffusive interactions, whereby such intermittent, recurrent collective bursting activity coexists with a completely synchronous state. An important ingredient of local dynamics is that it conforms to relaxation oscillations close to a canard transition \cite{VW15,WMR13,K15} between subthreshold and relaxation oscillations. Blending a recently introduced concept of phase-sensitive excitability of a periodic orbit \cite{FOW18,EFW19,Franovic2022a} and the interaction-induced trapping of orbits \cite{Medeiros2018,Medeiros2019,Medeiros2021}, we explain the mechanism by which the interplay of interactions and the vicinity of canard transition results in quenching of relaxation oscillations. This gives rise to patterns of rare spiking which under variation of coupling strength may self-organize into avalanche-like activity with scale-invariant features. We further show that avalanche patterns emerge in vicinity of a transition between two collective regimes with a lower and higher spiking rates, exhibiting classical indicators of criticality, such as a decreased resilience to perturbations and critical slowing down \cite{Meisel2015,Cocchi2017,Wilkat2019,Maturana2020}.

The paper is organized as follows: Section~\ref{sec:math_model} provides the details of the model and outlines the aspects of singular perturbation theory relevant to the explanation given in Sec.~\ref{sec:general_results} on how the interplay of interactions and structures associated with local multiple timescale dynamics may quench the spiking activity. In Sec.~\ref{sec:avalanches}, we investigate the statistical features of avalanche patterns, and show that these patterns emerge at the transition where the system displays classical criticality features in 
response to external stimulation. Section~\ref{sec:discussion} contains our concluding remarks and 
outlook.

\section{Array of non-locally coupled FitzHugh-Nagumo units} \label{sec:math_model}

\begin{figure}[!ht]
	\begin{center}
		\includegraphics[scale=0.35]{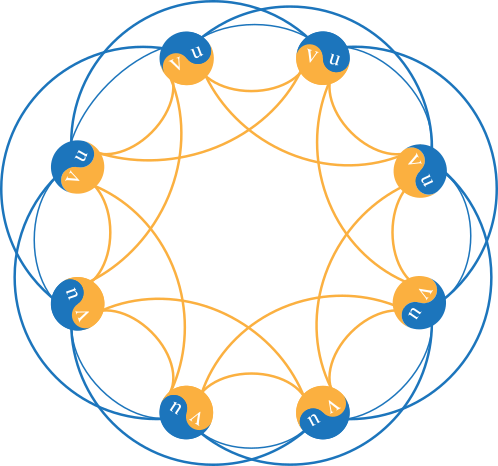}
		\caption{An array of FHN neurons for $N=8$ and $P=2$. Fast variables $u$ are represented in blue, while slow variables $v$ are shown in yellow.}
		\label{figure1}
	\end{center}
\end{figure}

Our model is an array of $N$ identical FHN units \cite{I07} with a simple non-local interaction scheme where each unit is coupled to $P$ of its neighbors to its left and to its right on a 1-dimensional ring:
\begin{eqnarray}
\nonumber
	\epsilon \dot{u}_{i} &=& u_{i} - \frac{u^{3}_{i}}{3} - v_{i}  + \frac{\sigma}{2P} \sum_{j=i-P}^{i+P} (u_{j}-u_{i}), \\
	\dot{v}_{i} &=& u_{i} + \alpha +  \frac{\sigma}{2P} \sum_{j=i-P}^{i+P} (v_{j}-v_{i}).
	\label{eqn:u_net}
\end{eqnarray}
All the indices are periodic modulo $N$. Due to the smallness of the parameter $\varepsilon \ll 1$,
here set to $\varepsilon=0.05$, the local dynamics feature a slow-fast structure with the fast
(activator) variables $u_i$ representing neuronal membrane potentials and the slow (recovery) variables
$v_i$ reproducing the coarse-grained behavior of ion-gating channels. The non-local interactions
are assumed to be linear (diffusive) and act between the activator/recovery variables in the units' fast/slow subsystems \cite{SF18,SFST18,SF17}, see the coupling scheme in Fig.~\ref{figure1}. Apart from the coupling radius $p=P/N$, the interactions are characterized by the coupling strength $\sigma$, and are considered to be attractive ($\sigma>0$) and homogeneous over the array.

Local dynamics is controlled by the bifurcation parameter $\alpha>0$, such that the singular Hopf bifurcation at $\alpha=1$ mediates the transition between the excitable regime ($\alpha \gtrapprox 1$), featuring a stable equilibrium $(u^*,v^*)=(-\alpha,-\alpha+\alpha^3/3)$, and the oscillatory regime ($\alpha < 1$) \cite{I07}. Within the framework of singular perturbation theory \cite{K15}, which treats the limit $\varepsilon \rightarrow 0$, an isolated FHN system has been shown to exhibit special type of trajectories, called \emph{canards}, that closely follow the repelling branch of the slow manifold for an appreciable time \cite{VW15,WMR13,K15} instead of rapidly departing from it. For small but finite $\varepsilon$, such trajectories form an exponentially thin layer, whereby there exists a so-called \emph{maximal canard} \cite{KS01} that follows the entire repelling branch of the slow manifold. The presence of such trajectories strongly impacts the behavior of the bifurcating limit cycle when decreasing $\alpha$ further below the bifurcation threshold $\alpha=1$. In particular, the incipient limit cycle undergoes a \emph{canard transition} \cite{DGKK12}, where its amplitude sharply increases within a narrow interval of $\alpha$ values exponentially small in $\varepsilon$. The canard transition mediates between small-amplitude harmonic oscillations of period $\mathcal{O}(\sqrt{\varepsilon})$ and large-amplitude relaxation oscillations of period $\mathcal{O}(1)$. In the language of neuroscience, this corresponds to a transition from subthreshold oscillations to the regime of tonic spiking. A classical result from asymptotic expansion theory is that the canard transition occurs at $\alpha=\alpha_c=1-\varepsilon/8$, cf. Ref.~\cite{BE86}.

Throughout this paper, the local bifurcation parameter is set to $\alpha=0.99$, the value below $\alpha_c$, such
that it supports relaxation oscillations. Nonetheless, the vicinity of the canard transition still influences the way the system responds to perturbations, be it due to interactions and/or noise. In particular, while the subthreshold oscillations below the canard transition manifest excitability in the classical sense \cite{MNV00}, it has recently been reported that the relaxation oscillations show a specific type of excitable behavior called phase-sensitive excitability of a limit cycle \cite{FOW18}. The latter comprises a non-uniform response to perturbations along the orbit of relaxation oscillations, such that the FHN system provides a nonlinear threshold-like response to perturbations during the passage close to the unstable equilibrium $(u^*,v^*)$. Then, perturbations of sufficient amplitude and acting in the appropriate direction are capable of inducing one or more subthreshold oscillations around the unstable equilibrium, whereby the maximal canard acts as the threshold manifold. The emergence of such subthreshold oscillations in response to interactions will later prove important for understanding the mechanism giving rise to nontrivial collective dynamics behind the activity avalanches.

Our primary interest concerns the impact of coupling strength $\sigma$ on the system's dynamics, focusing
on the case of weak interactions $\sigma \in [0,0.1]$. All the numerical experiments have been performed for the system size $N=50$ and coupling range $P=10$ unless stated otherwise. The numerical integration has been performed using the Cash-Karp (4, 5) method with adaptive stepsize control implemented via GNU Scientific Library (GSL) \cite{GSL21}. The time series in the remainder of the paper illustrate the  asymptotic system behavior after discarding a sufficiently long transient of, e.g., $5 \times 10^3$ time units. When illustrating the dependence on $\sigma$, the coupling strength increment is $\Delta \sigma=10^{-3}$. For each value of $\sigma$, we consider a set of 10 different random initial conditions $(\vec{u}_{0},\vec{v}_{0}) \in [-2,2]^N \times [-2,2]^N$ . 

We find that a range of coupling strengths supports the onset of a regime where irregular asynchronous rare spiking activity is interspersed with brief intervals of cascading pseudo-synchronous bursting activity, called avalanches. The described regime is bistable with the regime of synchronous regular spiking activity, as we will demonstrate in Sec.~\ref{sec:general_results}.

\section{Array dynamics in dependence of coupling strength} \label{sec:general_results}

Given that the units are identical and interact by attractive diffusive couplings, the system~\eqref{eqn:u_net} possesses an invariant synchronization manifold $u_1(t)=u_2(t)=...=u_N(t), v_1(t)=v_2(t)=...=v_N(t)$. Since the isolated dynamics of neurons comprises relaxation oscillations, this manifold contains a limit cycle attractor where all the units perform identical relaxation oscillations. 
In the following, we will show that under variation of the coupling strength $\sigma$, the system~\eqref{eqn:u_net} may exhibit non-trivial emergent dynamics that unfolds off the invariant synchronization manifold. In other words, we find a range of $\sigma$ values where due to non-local interactions, not all of the initial conditions converge to the invariant manifold, and the completely synchronized relaxation oscillations coexist with another type of collective dynamics.

To observe such emergent dynamics, we introduce a global order parameter $\mu$ that characterizes the synchronization of units' \emph{average spiking frequencies}. Unlike the more classical synchronization parameters, involving synchronization error or average local variances from the mean variables, $\mu$ is not indented to quantify both frequency and phase synchronization of units, but rather to describe the quenching of units' average spiking frequencies due to non-local interactions. By construction, $\mu$ is introduced to indicate the relative persistence of units' trajectories in the neighborhood of the limit cycle $S$ corresponding to relaxation oscillations of an \emph{uncoupled} (isolated) unit. To define $\mu$, we first denote by $K$ the spike count of an uncoupled unit within a sufficiently long time interval $\Delta T$. Then, for the system of coupled units \eqref{eqn:u_net}, we consider $J_i$ as the spike count of a unit $i$ within the time interval $\Delta T$. Using these two quantities, the global order parameter $\mu$ is given by 
\begin{align}
 \mu=\frac{1}{NK}\sum\limits_{i=1}^N J_i.  \label{eq:mu}
\end{align}
Qualitatively, $\mu$ compares the ensemble-averaged spiking frequency of coupled units to the spiking frequency of an uncoupled unit. Naturally, these two frequencies are equal, resulting in $\mu=1$, when the system's state lies on the large-amplitude limit cycle on the invariant synchronization manifold. Nevertheless, note that $\mu=1$ also corresponds to such states where the units are not on the synchronization manifold, but perform relaxation oscillations mutually shifted in phase. One expects the emergent dynamics with quenched spiking of individual units to be characterized by $\mu<1$.

\begin{figure}[!ht]
	\begin{center}
		\includegraphics[scale=0.48]{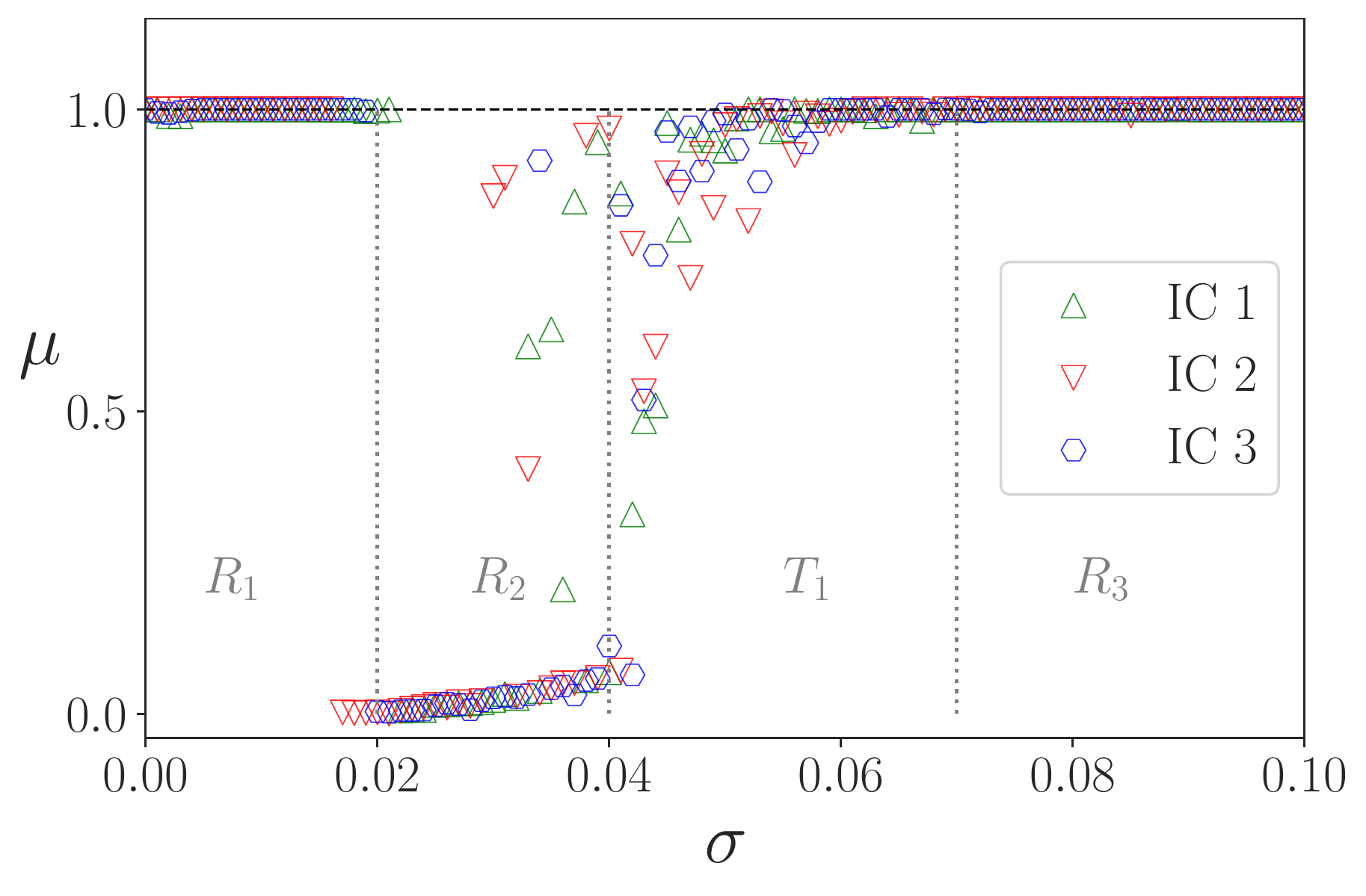}
		\caption{Order parameter $\mu$ in dependence of coupling strength $\sigma$ for three different sets of initial conditions (green up-triangles, red down-triangles, and blue circles). Intervals of $\sigma$ denoted by $R_1$ and $R_3$ are characterized by the prevalence of relaxation oscillations ($\mu \approx 1$). Intervals $R_2$ and $T_1$ respectively support coexistence of asynchronous states $\mu \approx 0$ and $0<\mu<1$ with the completely synchronous state. The dashed line $\mu=1$ corresponds to the case where all the units perform completely synchronous relaxation oscillations.}
		\label{figure2}
	\end{center}
\end{figure}

Figure~\ref{figure2} shows the order parameter $\mu$ in terms of the coupling strength $\sigma$ for asymptotic dynamics obtained from three different sets of initial conditions (green up-triangles, red down-triangles, and blue circles) in the weak coupling regime ($\sigma \ll 1$). For sufficiently small $\sigma$, all the initial conditions lead to frequency synchronized relaxation oscillations of individual units, see the region $R_1$. Nevertheless, when increasing $\sigma$, one observes an interval $\sigma \sim [0.02,0.07]$ that supports asynchronous states characterized by the global order parameter $\mu<1$. Such states emerge only for certain sets of initial conditions, and the synchronous state coexists throughout the entire $\sigma$ interval. By the corresponding values of $\mu$, one may distinguish between two types of asynchronous states:  (i) the region $R_2$ where the global order parameter attains very small values $\mu \approx 0$; (ii) an interval $T_1$ where $\mu$ values of asynchronous states are enhanced but still lie notably below the $\mu=1$ level. For a stronger coupling strength $\sigma$, one finds region $R_3$ where the synchronous state is regained for all sets of initial conditions. The same physical picture qualitatively holds for a range of coupling radii $p$. Nevertheless, the width of the interval of intermediate $\sigma$ values supporting asynchronous states reduces with $p$, eventually vanishing for interactions of sufficiently long-range.

\begin{figure}[!ht]
\centering
\includegraphics[scale=0.53]{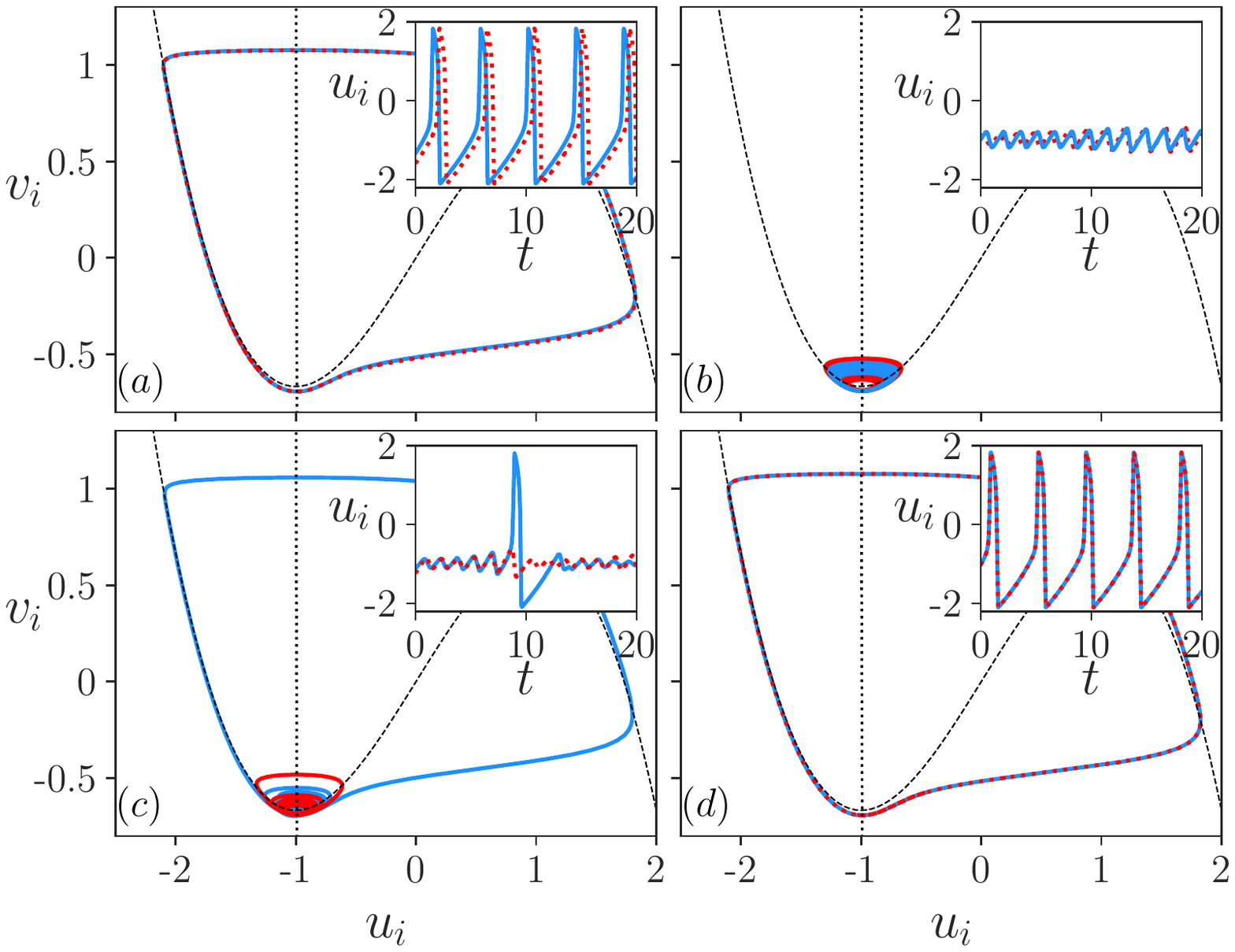}
\caption{Main frames: orbits $(u_i(t),v_i(t))$ of two units $i=20$ (blue solid lines) and $i=50$ (red dashed lines); insets: time traces $u_i(t)$ of the same two units for different system states. (a) $\sigma=0.001$: phase-shifted synchronization of relaxation oscillations. (b) $\sigma=0.02$: the feedback from local mean-fields causes trapping of orbits around the unstable equilibrium $(u^{*},v^{*})$. (c) $\sigma=0.041$: the orbits eventually escape from the trapping region, generating rare spikes. (d) $\sigma=0.1$: coupling strength is sufficient to induce complete synchronization of relaxation oscillations. The unstable equilibrium $(u^{*},v^{*})$ of isolated dynamics lies at the intersection of nullclines (black dashed and dotted curves).}\label{figure3}
\end{figure}

To gain a deeper insight into the emergent dynamics typical for different $\sigma$ intervals, we consider the corresponding state-space projections $(u_i, v_i)$ for two representative units, indicated in blue solid and red dashed lines in Fig.~\ref{figure3}. For $\sigma = 0.001$, which lies in the region $R_1$, the neurons already perform relaxation oscillations along the same orbit but are shifted in phase, cf. the orbits and the
time traces in Fig.~\ref{figure3}(a). For this small coupling strength, the phases remain free along the limit cycles at the state
space of different FHN units\cite{Pikovsky2001}. Within the region $R_2$, represented by $\sigma = 0.02$ in Fig.~\ref{figure3}(b), the units mostly
perform small-amplitude oscillations around the unstable equilibrium $(u^*, v^*)$, and only a few or none of the units occasionally escape the trapping region generating rare spikes. Trapping of the trajectories in the vicinity of the unstable equilibrium derives from the impact of
local mean-fields, whose fluctuations are reflected in the amplitude variability of subthreshold oscillations around $(u^*, v^*)$. The localized excitations (spikes) become more prevalent for a larger $\sigma = 0.041$ that belongs to the interval $T_1$, see Fig.~\ref{figure3}(c). Increasing $\sigma$ within $T_1$, one observes patterns comprised of local mixed-mode oscillations \cite{DGKK12} where units fire more frequently and more correlated. The statistical properties of such solutions are a key aspect of this study and will be elucidated in the next sections. Finally, for $\sigma = 0.1$ from the region $R_3$, the system dynamics are characterized by completely (both frequency and phase) synchronized relaxation oscillations of individual units, cf. Fig.~\ref{figure3}(d).

Now let us focus on the mechanism causing the trapping of units' orbits in the vicinity of the unstable fixed point $(u^{*},v^{*})$. First, we recall the notion of phase-sensitive excitability of a periodic orbit invoked in Sec.~\ref{sec:math_model}. At variance with \cite{FOW18}, which introduced this notion while analysing the non-uniform response of relaxation oscillations to noise, a similar type of effect emerges due to non-local interactions. Namely, the units whose isolated dynamics comprise relaxation oscillations become trapped and perform subthreshold oscillations around the unstable equilibrium. Then, the maximal canard establishes a state space threshold separating the transient small-amplitude oscillation from the limit cycle of relaxation oscillations. For deterministic networked systems, the trapping of trajectories has previously been observed in the vicinity of more complex invariant sets. In particular, in Ref.\cite{Medeiros2018,Medeiros2019,Medeiros2021}, it has been demonstrated that the couplings can trap units' trajectories in the vicinity of unstable chaotic sets. Then, the trapping mechanism is based on an interplay between interactions and the dynamics in the chaotic set, which creates random perturbations that prevent the trajectories from escaping the vicinity of the invariant set via its unstable manifold. The chaotic sets involved in the trapping occur in the state space of each unit. The latter is similar to the scenario here, but instead of chaotic sets, we consider the trapping mediated by unstable equilibria encircled by the maximal canards.

\begin{figure}[!ht]
\centering
\includegraphics[scale=0.42]{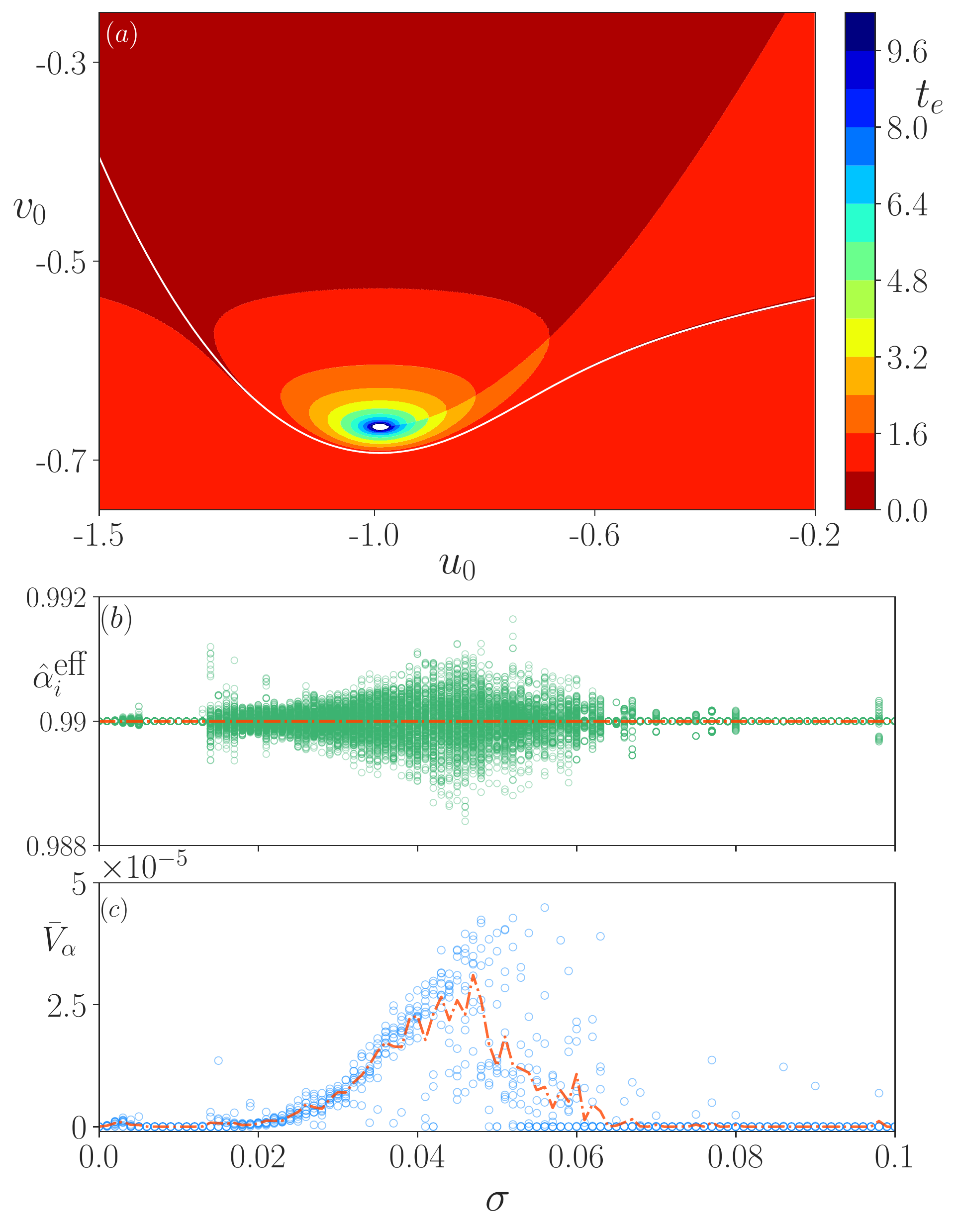}
\caption{(a) For an isolated FHN unit, i.e., Eq.~\eqref{eqn:u_net} with $\sigma=0$, the color scheme indicates escape times $t_e$ from the maximal canard region for different initial conditions $(u_0,v_0)$; white curve: segment of the limit-cycle $S$ close to the maximal canard. (b) Time-averaged effective bifurcation parameters $\hat{\alpha}_{i}^\text{eff}$ as a function of $\sigma$. (c) Blue dots: local variances $V_{\alpha,i}$ of effective bifurcation parameters $\alpha_{i}^\text{eff}(t)$; red dash-dotted curves in (b) and (c): population-averaged values for different $\sigma$.}
\label{figure4}
\end{figure}

In the following, we propose a mechanism to explain the dynamics in the intervals $R_2$ and $T_1$ that contains the main ingredients of both phase-sensitive excitability of periodic orbits and the above described trapping phenomenon. We begin by revisiting the dynamics of an isolated FHN neuron where the maximal canard provides a threshold between different types of orbits, distinguished by the motion around the unstable fixed point $(u^{*},v^{*})$. The differences between the associated transients become apparent if one determines the corresponding {\it escapes time} $t_e$ from the region enclosed by the maximal canard. This quantity expresses the dimensionless time required for trajectories starting from different initial conditions to reach the limit cycle of relaxation oscillations $S$. In Fig.~\ref{figure4}(a), color coding indicates the escape times $t_e$ for a large set of initial conditions $(u_0,v_0)$. Note the thin boundaries between the regions with different values of $t_e$ that reflect the spiraling of the maximal canard around $(u^{*},v^{*})$, and the white line just below indicating a segment of the orbit of the limit cycle corresponding to relaxation oscillations. The subtlety of such boundaries makes the system highly sensitive to perturbations. For instance, a trajectory in the maximal canard region with a certain prescribed escape time, if perturbed, may change its current escape route and perform extra loops around $(u^{*},v^{*})$. The same applies to the orbit of relaxation oscillations, which under the effect of an appropriate perturbation, may be injected into the maximal canard region when passing close to it so to perform loops around the unstable fixed point.

Let us now focus on the case of FHN neurons embedded in an array. There, it is the non-local interactions that provide perturbations to local dynamics, sensitively affecting the units' orbits around the maximal canard. Depending on the character of perturbations, the trajectories of only a subset of neurons may undergo subthreshold oscillations due to trapping by the maximal canard, while the remaining neurons continue to perform relaxation oscillations. 
Such a scenario gives rise to an emergent asynchronous behavior. Since the coupling function is diffusive, its amplitude increases in a desynchronized network, contributing to larger perturbations to neuronal dynamics. Consequently, the interaction between the perturbation-sensitive dynamics around the maximal canard and the couplings, i.e., local mean-fields, constitutes a positive feedback loop. One may numerically assess the range of coupling strengths $\sigma$ where such an impact of interactions is the strongest. Appreciating that the interactions introduce a parametric perturbation of local neuronal dynamics, we introduce an effective bifurcation parameter $\alpha_i$ for each neuron as
\begin{equation}\label{eqn:a_eff}
	\alpha_{i}^\text{eff}(t) = \alpha +  \frac{\sigma}{2P} \sum_{j=i-P}^{i+P} (v_{j}(t)-v_{i}(t)),
\end{equation}
where $\alpha = 0.99$ is the unperturbed value defined in Sec.~\ref{sec:math_model}.

Figure~\ref{figure4}(b) depicts the time averages $\hat{\alpha}_{i}^\text{eff}$ of the effective parameter $\alpha_{i}^\text{eff}(t)$ as a function of $\sigma$. One observes that for $\sigma\lesssim 0.018$, the value of $\hat{\alpha}_{i}^\text{eff}\approx 0.99$ approximately equals that of an isolated unit. Here, the amplitude of  perturbations from the local mean-fields is subthreshold and cannot induce small-amplitude oscillations around $(u^{*},v^{*})$. Consequently, all the units perform relaxation oscillations, cf. region $R_1$ in Fig.~\ref{figure2}. However, for $\sigma \approx 0.018$, the couplings become capable of  trapping the units within the canard region to generate small-amplitude oscillations. In parallel, one observes that the value of $\hat{\alpha}_{i}^\text{eff}$ begin to substantially depart from the unperturbed value $\alpha=0.99$, cf. Fig.~\ref{figure4}(b). Such increasing deviations are associated with the feedback from non-local interactions, whose impact on system dynamics grows as the desynchronization sets in. Enhancing $\sigma$ further, the contribution from non-local interactions to $\hat{\alpha}_{i}^\text{eff}$ peaks around $\sigma\approx 0.04$. There, the parametric perturbation to units shows a high variability over the array, cf. the increase in the corresponding variances $V_{\alpha,i}$ of effective bifurcation parameters in Fig.~\ref{figure4}(c). The given value of $\sigma$ approximately corresponds to the transition between regions $R_2$ and $T_1$ from Fig.~\ref{figure2}. As $\sigma$ is further increased, the attractive nature of the couplings begins to dominate the dynamics, contributing to units' synchronization. This is accompanied by the decrease of the amplitudes of parametric perturbations affecting the units up to the point where they become subthreshold, such that the units again perform relaxation oscillations, cf. region $R_3$ in Fig.~\ref{figure2}.

In Sec.~\ref{sec:avalanches}, we will explore the statistical properties of the network solutions. We pay special attention to the transition between $R_2$ and $T_1$, where the non-local coupling and sensitive response to perturbations of relaxation oscillations in the vicinity of the maximal canard make the largest impact.

\section{Avalanche activity} \label{sec:avalanches}

As elaborated in Sec.~\ref{sec:general_results}, for a range of intermediate $\sigma$ in Fig.~\ref{figure2}, one
finds activity patterns where the units spend much time trapped by the maximal canard in the vicinity of 
the unstable fixed point $(u^*,v^*)$, while being rarely released to perform spikes. In the following, we 
resolve the spatio-temporal structure of such emergent states showing that they conform to an avalanche-like 
activity, where intermittent pseudo-synchronous spiking, localized to various degrees, is separated by long 
periods of quiescence over the array. Note that the observed avalanches are not intended to model classical neuronal avalanches \cite{Beggs2003,Beggs2004,Plenz2007,Hahn2010}, though a partial analogy may be drawn, as discussed in
Sec.~\ref{statistics}.

\begin{figure*}[t]
\centering
\includegraphics[scale=0.75]{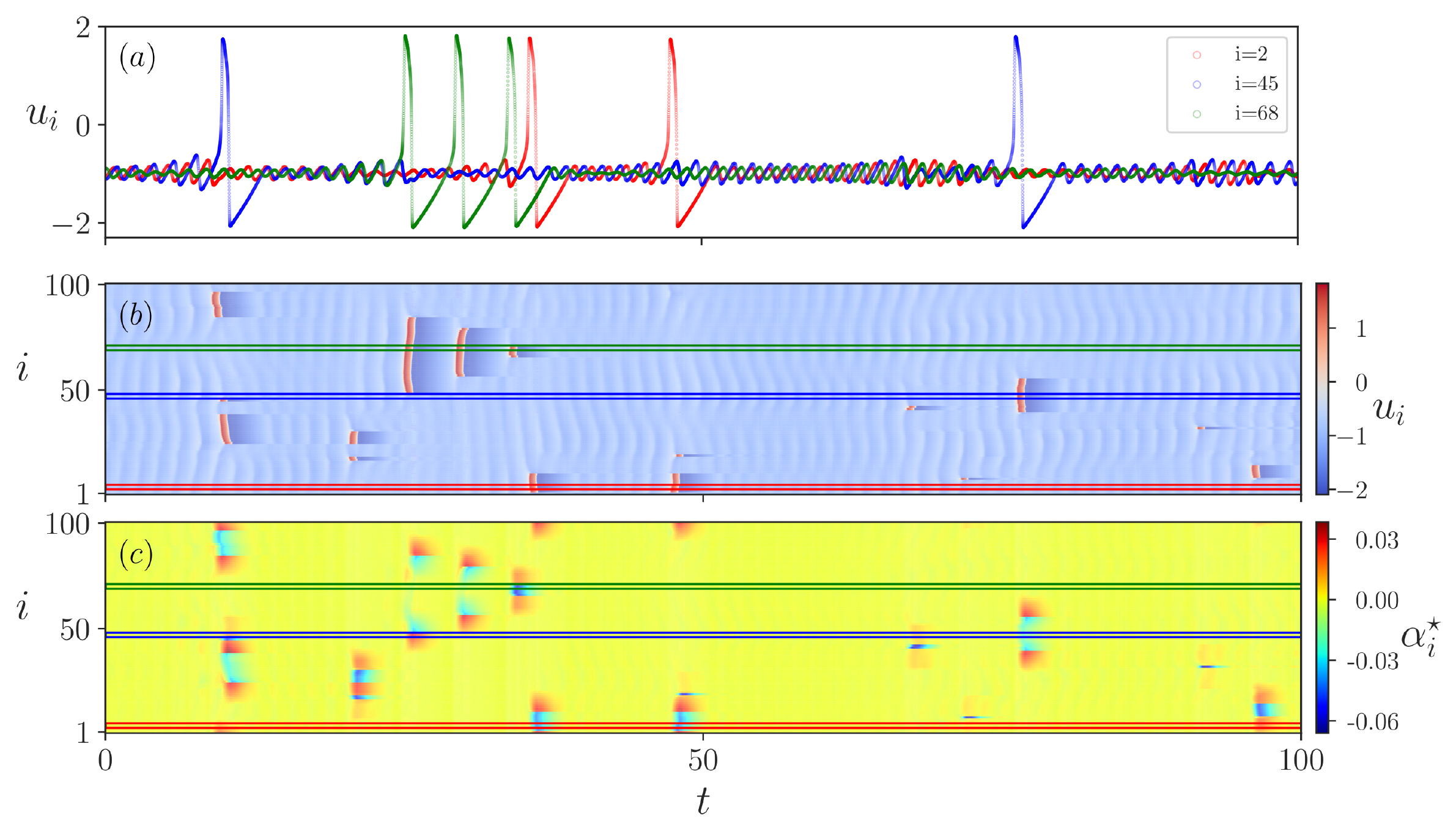}
\caption{Self-organization of avalanches. (a) Time traces $u_i(t)$ for three units $i=2$, $i=45$, and $i=68$, indicated by green, blue, and red rectangles in panel (b), respectively. (b) Spatio-temporal evolution of fast variables $u_i(t)$. (c) Spatio-temporal evolution of the quantity $\alpha_{i}^*(t)=\alpha_i^\text{eff}(t)-\alpha_c$, which shows that the units spend most of the time in the vicinity of the canard transition. System parameters: $\sigma=0.04$, $N=100$ and $p=0.2$.}
\label{figure5}
\end{figure*}

Let us first consider the spatio-temporal evolution of local membrane potentials $u_i(t)$ described by Eq.~\eqref{eqn:u_net}, see Figs.~\ref{figure5}(a) and \ref{figure5}(b). Indeed, the latter indicates that the typical activity patterns are self-organized into episodes of pseudo-synchronous spiking separated by silent episodes. Nevertheless, in terms of temporal organization, two types of avalanches may be distinguished, namely \emph{cascading} 
events, cf. the example of an avalanche beginning around $t\approx 30$ in Fig.~\ref{figure5}(b), where the (spatially localized) spiking activity propagates forming temporal sequences; and \emph{temporally localized} (isolated) events, where the (spatially localized) spiking occurs within a narrow time window. Note that the duration of the time window used to identify pseudo-synchronous spiking is specified in Sec.~\ref{statistics}. Qualitatively, the episodes of spiking activity resemble self-localized excitations in excitable media \cite{Wolfrum2015,Franovic2022b}. The cascading events have a step-pyramid-like space-time structure. This reflects the fact that at every next level, only the units closer to the center of the previous level perform a spike. The latter units remain active because they receive most of the input from the spiking rather than the silent units. Naturally, the units at the top level, e.g., unit $i=68$ in Fig.~\ref{figure5}(a) and Fig.~\ref{figure5}(b), fire more spikes during a cascading event than the units whose spiking terminates at some of the lower levels. In contrast to cascading avalanches, each unit participating in a temporally localized event spikes only once. In terms of spatial organization, the units spiking within a given narrow time window can appear as connected clusters, or may display a multi-cluster structure forming spatially disconnected clusters.

The intrinsic structure and self-organization of avalanche patterns can be described in more detail by looking into the spatio-temporal evolution of the quantity $\alpha_i^{*}(t)=\alpha_i^\text{eff}(t)-\alpha_c$ as shown in Fig.~\ref{figure5}(c). In particular, one observes that units spend most of the time in vicinity of the canard transition $\alpha_i^*\approx 0$, which underlies the important role of the canard transition in the self-organization of avalanches. Moreover, for the cascading events, one observes that the units around the excited region are furthest above the bifurcation threshold, i.e., have the largest values of $\alpha_i^*(t)$. This effectively facilitates the pattern confinement, making the avalanche events spatially localized.  

\begin{figure}[!ht]
	\centering
	\includegraphics[scale=0.34]{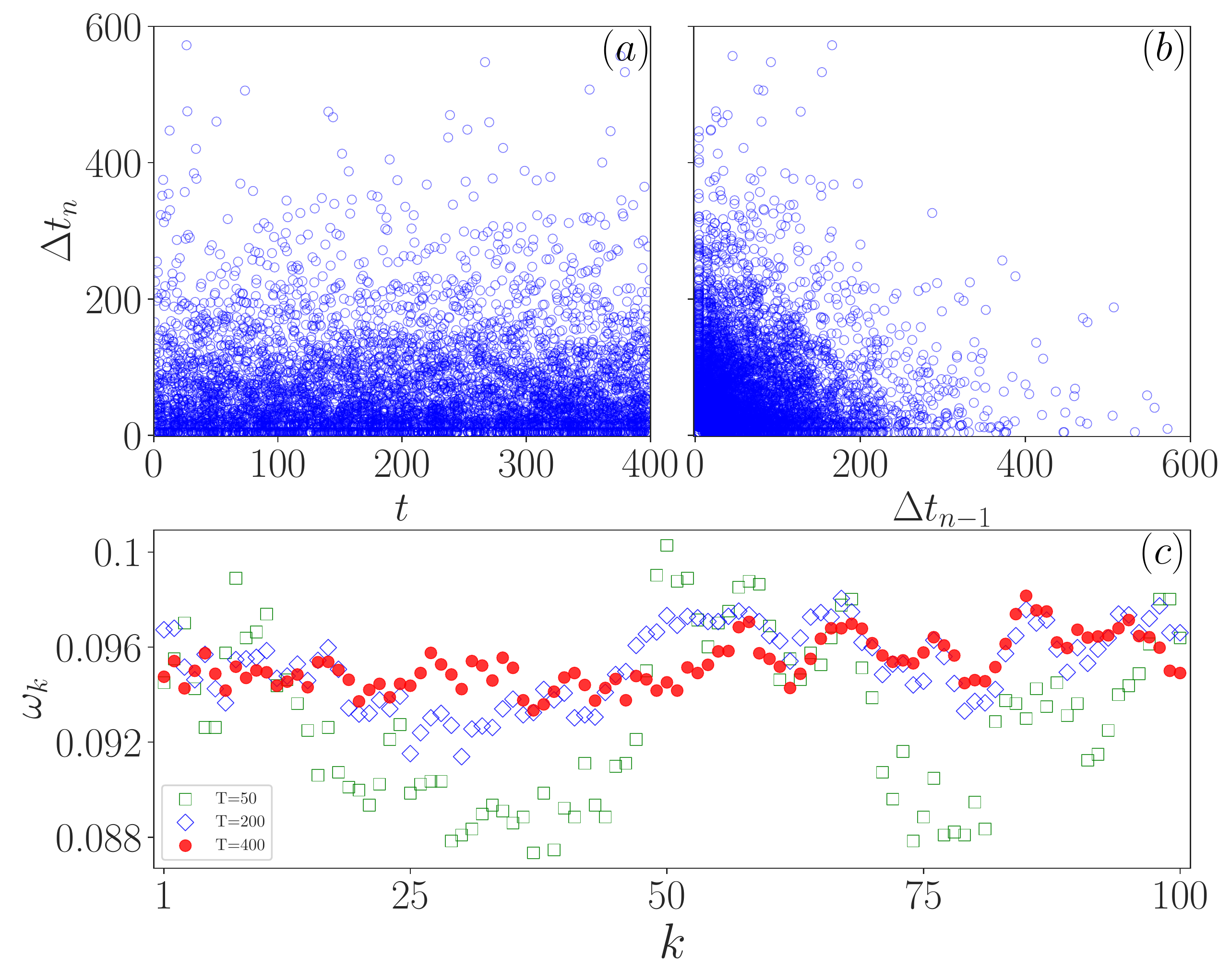}
	\caption{(a) Temporal evolution of the return times $\Delta t_n(t)$ to the Poincar\'{e} cross-section $u_k(t)=1, \dot{u}_k(t)>0$ of a single unit. (b) First return map $\Delta t_{n}(\Delta t_{n-1})$ of successive return times to the Poincar\'{e} cross-section. (c) Spatial distribution of average spiking frequencies $\omega_k$ over time periods $T=5 \times 10^{4}$ (empty squares), $T=2 \times 10^{5}$  (empty diamonds) and $T= 4 \times 10^{5}$ (solid circles). System parameters: $\sigma=0.04$, $N=100$, $p=0.2$.}
	\label{figure6}
\end{figure}

The local dynamics comprise irregular mixed-mode oscillations, involving fast subthreshold oscillations interspersed with random rare spikes, cf. Fig.~\ref{figure5}(a) which illustrates the time traces $u_i(t)$ of three units highlighted in Fig.~\ref{figure5}(b). The irregularity of single units' interspike intervals is corroborated by Fig.~\ref{figure6}(a) showing the temporal evolution of the return times $\Delta t_n(t)$ to the Poincar\'{e} cross-section $u_k(t)=1, \dot{u}_k(t)>0$ for an arbitrary unit. Together with the corresponding first return map of successive return times $\Delta t_{n}(\Delta t_{n-1})$ in Fig.~\ref{figure6}(b), it evinces that the units may sometimes fire spikes in close succession, but that there may also be long periods of quiescence. The spatial profile of average spiking frequencies $\omega_k=2\pi M_k/T$, where $M_k$ is the spike count within a macroscopic time interval $T$, is shown in Fig.~\ref{figure6}(c). Expectedly, as the averaging time interval is increased, the $\omega_i$ profile becomes more uniform, indicating that it should appear flat for very long $T$ as the spiking excitations occur randomly in space. Qualitatively, our scenario involving rare and irregular recurrent spiking bears certain resemblance to the onset of extreme events in systems of diffusively coupled nonidentical FHN units with excitable local dynamics \cite{ansmann2013,Karnatak2014,Bialonski2015,Ansmann2016}, as well as identical FHN oscillators with delayed diffusive couplings \cite{Saha2017,Saha2018}. However, in contrast to \cite{ansmann2013}, we typically find spatially localized events, rather than the bursting events spanning the entire network.

\subsection{Statistical features of avalanches}\label{statistics}

In this Section, our goal is to address in detail how the statistical features of activity patterns like the one in Fig.~\ref{figure5}(a) depend on the coupling strength $\sigma$. Let us first precisely define the avalanche events and the associated properties we are interested in. Starting from a set of random initial conditions $(\vec{u}_{0},\vec{v}_{0})\in [-2,2]^N \times [-2,2]^N$, we consider the evolution of an array Eq.~\eqref{eqn:u_net} over the interval $\Delta T =5 \times 10^4$.
An individual avalanche event comprises a joint spiking activity of a \emph{cluster} of a certain number of units $k$ within the narrow time window $\Delta t = 100 \delta t$, where $\delta t$ is the integration step. The avalanche size, denoted by $S_k$, then refers to the number of units that have fired at least once during this small interval, and is not related to the total number of spikes emitted by the units forming the cluster. In other words, $S_1$ denotes an event where only a single unit has fired within the given time window $\Delta t$, whereas $S_{N}$ corresponds to an avalanche spanning the whole array. To elucidate how the avalanche properties depend on $\sigma$ without a potential bias due to initial conditions, for each value of $\sigma$ we perform numerical experiments with 10 different sets of random initial conditions.

\begin{figure}[!htp]
	\begin{center}
		\includegraphics[scale=0.6]{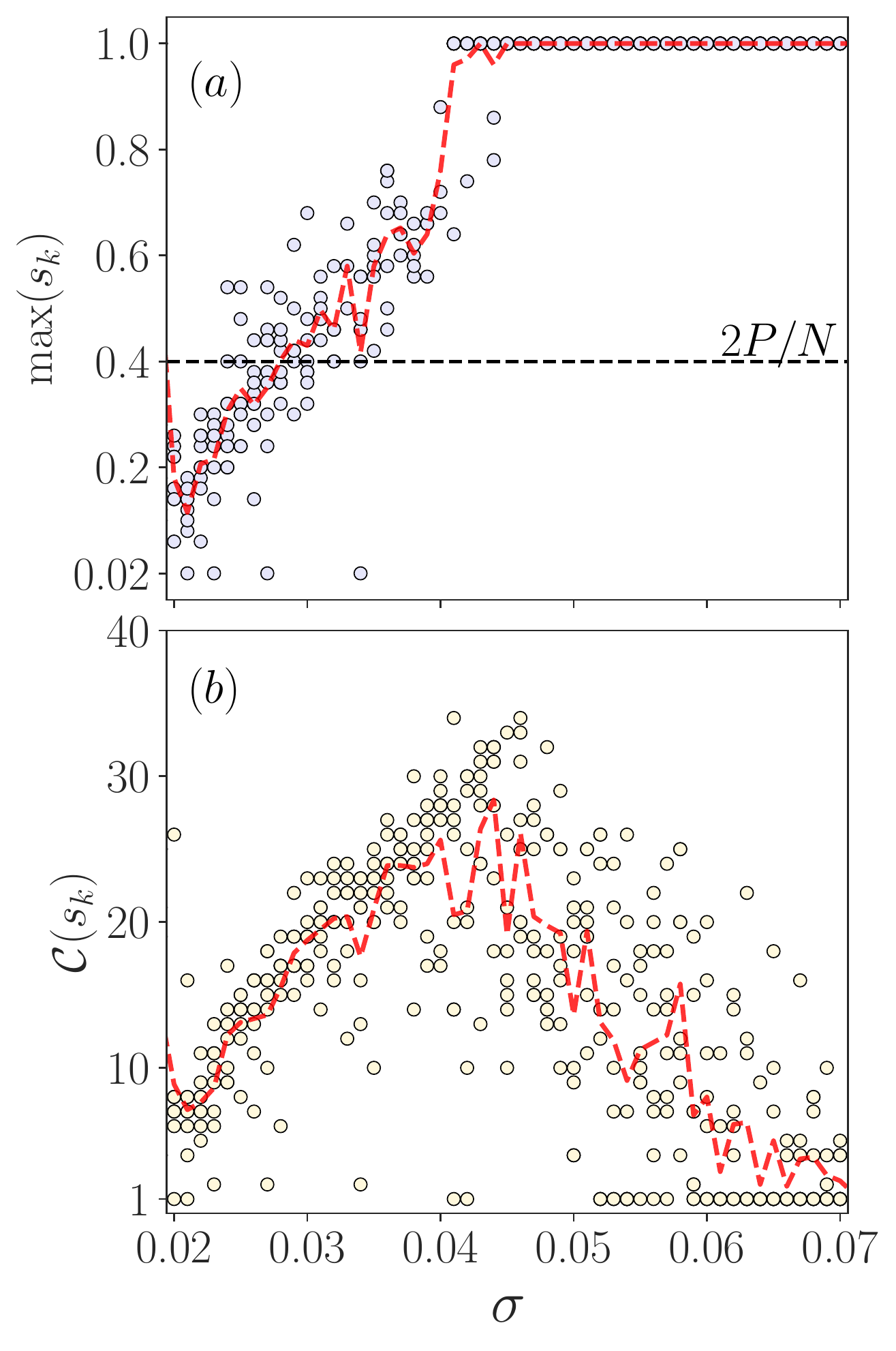}
		\caption{Statistical properties of avalanches. (a)  Largest, relative avalanche sizes $\max(s_{k})$ in terms of $\sigma$. For each $\sigma$, dots indicate the results for 10 different initial conditions. The average values (red curve) exceed the connectivity $2p=2P/N$ (green dashed line) for $\sigma>0.03$. (b) Number of different cluster sizes $\mathcal{C}(s_{k})$ as a function of $\sigma$. The average (red curve) shows a peak in the vicinity of the transition between regions $R_2$ and $T_1$, cf. Figs.~\ref{figure2} and \ref{figure4}.}
		\label{figure7}
	\end{center}
\end{figure}

Focusing on the $\sigma$ interval associated with regimes $R_2$ and $T_1$, Fig.~\ref{figure7}(a) illustrates the $\sigma$ dependence of the \emph{maximal} avalanche sizes $\max(s_{k})$ normalized
over the array size $N$, i.e., $s_k=S_k/N$. Multiple symbols for a given value of $\sigma$ denote the results obtained for the different sets of initial conditions, and the red curve indicates the values averaged over the ensemble of initial conditions. For smaller $\sigma$, even the maximal avalanche sizes do not exceed the normalized coupling range $2p=2P/N$, indicated by the horizontal green line. This implies that avalanches remain localized events focused around the initial excitation, or put differently, that the correlation length
of spontaneous activity fluctuations remains short. However, for larger coupling strengths
$\sigma \gtrapprox 0.025$, the average values over different initial conditions exceed the coupling range, suggesting that the synchronous spiking activity typically propagates over the array, indicating an increase in the system's correlation length. Enhancing the coupling strength further into the $T_1$ regime ($\sigma>0.04$), we observe that maximal avalanches indeed span the entire array. 

To get an insight into the variability of avalanche cluster sizes, in Fig.~\ref{figure7}(b) we show how the maximal number of different cluster sizes $\mathcal{C}(s_k)$ depends on $\sigma$. Multiple symbols for any given $\sigma$ again correspond to results for different
initial conditions. One observes that the variability of cluster sizes, reflected in the number of different recorded cluster sizes, reaches a maximum around $\sigma \approx 0.04$, the values near the transition between the regimes $R_2$ and $T_1$ from Fig.~\ref{figure2}. Nonetheless, within the $T_1$ regime, another form of variability increases. Namely, the diversity of cluster sizes recorded in simulations starting from different initial conditions becomes much more pronounced than in the $R_2$ regime.

Both the onset of avalanches that span the entire array in Fig.~\ref{figure7}(a), and the highest variability of avalanche sizes observed in Fig.~\ref{figure7}(b) for $\sigma \approx 0.04$, suggest that the change of regimes from $R_2$ to $T_1$ under increasing $\sigma$ bears signatures of criticality. One may draw a partial analogy to observations on resting state (spontaneous) activity in neuronal systems. There, the neuronal avalanches \cite{Beggs2003,Beggs2004,Plenz2007,Hahn2010}, found in electrophysiological recordings, both under in vitro and in vivo conditions, as well as by electroencephalography and functional magnetic resonance imaging, are known to show criticality features. Manifestations of criticality classically involve scale invariance in the distributions of relevant quantities, e.g. the size and duration of neuronal avalanches, which is reflected in the power-law behaviors of the form $F(x)\propto x^{-\gamma}$, where $\gamma$ is a critical exponent \cite{Nishimori2010,Kello2010}. Criticality features are generally associated with proximity to critical/phase transitions between ordered and disordered phases 
\cite{Chialvo2010,Plenz2014,Munoz2018,Wilting2019}, or in case of neuronal avalanches, between an absorbing state with a quickly decaying spiking activity and an active state with a runaway (exploding) activity propagation \cite{Scarpetta2013}. Nevertheless, the concept of phase transitions applies to systems in the thermodynamic limit $N\rightarrow \infty$, so an observation of genuine power-laws cannot be expected for finite-size systems. To resolve this, one often invokes the point that the phase transitions in finite systems extend over a critical region called Griffiths phase \cite{Moretti2013,Zimmern2020,GUT21}.
There, the system is quasi-critical and maintains certain aspects of criticality, including the truncated power-law behaviors (power laws with exponential cut-offs) of relevant quantities \cite{Yu2014}. This also applies to neuronal avalanches, where the classically reported exponents for the avalanche size and duration are $3/2$ (with some exceptions) and $2$, respectively, while the cut-off typically matches the system size \cite{Beggs2003,Shriki2013}, but may also deviate from it \cite{Klaus2011,Bellay2015}. One should further note that the power-law distributions of event sizes per se may not necessarily imply that the system is poised close to criticality \cite{Touboul2010,Touboul2017}. Conversely, there are instances, such as certain models of neuronal avalanches, where a critical system shows a scale-free distribution of event sizes that does not conform to a power-law \cite{Taylor2013}. Such results may partly derive from a potentially fuzzy relationship between the definition of observed events and the local dynamics behind them. 

\begin{figure}[!htp]
	\begin{center}
		\includegraphics[scale=0.46]{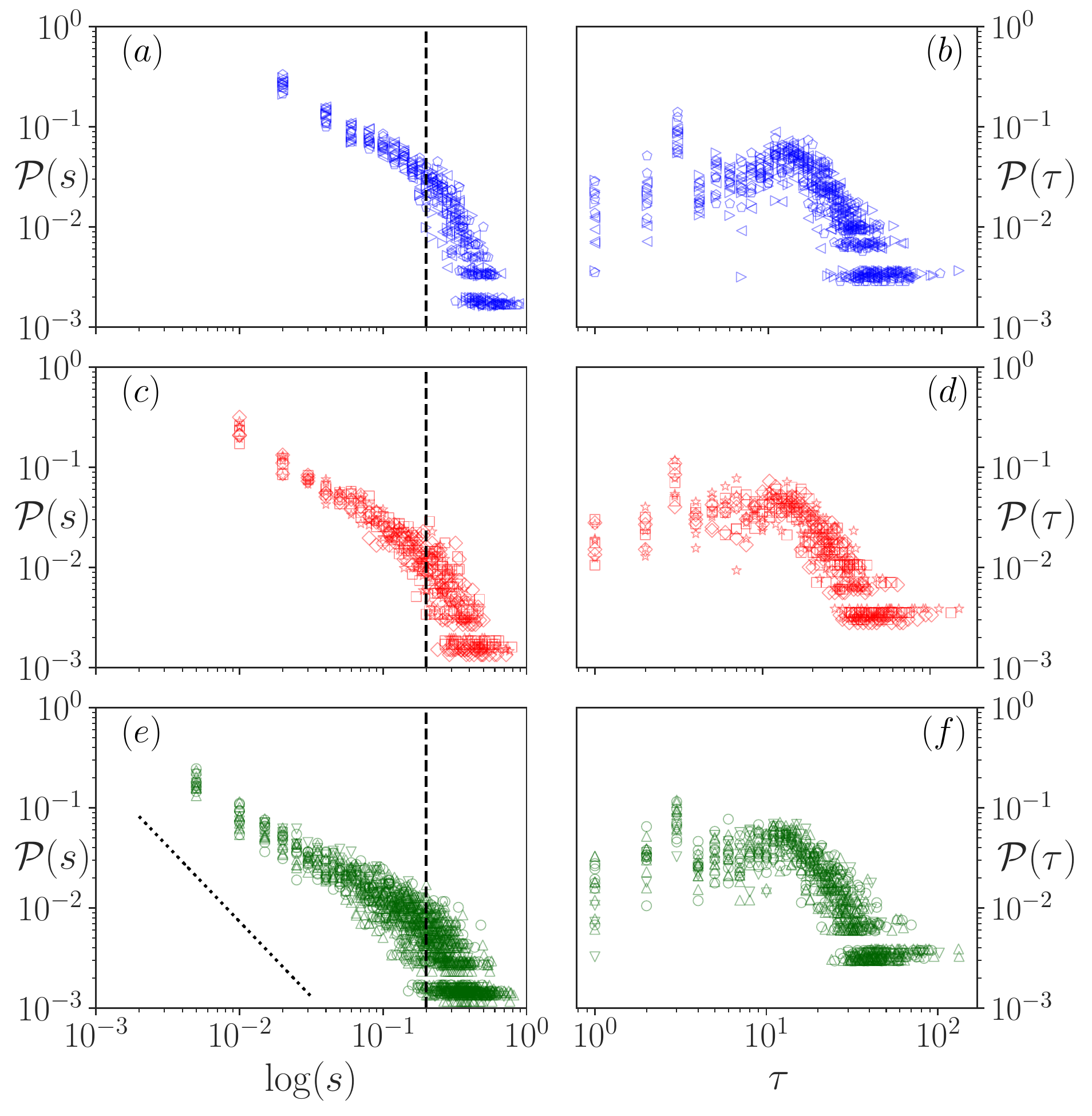}
		\caption{Distributions of relative avalanche sizes $s=S/N$ and laminar times $\tau$ for (a)-(b) $N=50$, (c)-(d) $N=100$ and (e)-(f) $N=200$. Different symbols indicate the results obtained for different sets of initial conditions. $\sigma$ is chosen in the vicinity of the transition between the regions $R_2$ and $T_1$. Coupling radius $p=P/N=0.2$ (vertical dashed black lines in the left column) is kept fixed in all the simulations. Distributions $\mathcal{P}(s)$ in (a), (c) and (e) show a power-law behavior for small and intermediate avalanches followed by a cut-off. The comparison with the power-law $\beta=3/2$ (black dotted line) in panel (e) is provided as a guideline. Distributions of laminar times $\mathcal{P}(\tau)$ in (b), (d) and (f) show a peak indicating the presence of a characteristic timescale.}\label{figure8}
	\end{center}
\end{figure}

Given the arguments above, we focus on the properties of avalanches in the narrow range
$\sigma \in [0.037,0.043]$ around the transition between the regimes $R_2$ and $T_1$ from Fig.~\ref{figure2}. In particular, fixing $\sigma$, we consider the probability distribution $\mathcal P(s)$ of relative avalanche cluster sizes $s=S/N$ and the probability distribution $\mathcal{P}(\tau)$ of time intervals $\tau$ between the successive avalanches, see the left and the right column in Fig.~\ref{figure8}, respectively. Both $\mathcal{P}(s)$ and $\mathcal{P}(\tau)$ are sampled for three different array sizes ($N=50$, $N=100$ and $N=200$) maintaining the fixed coupling radius $p=P/N=0.2$. For all three $N$ values, the distributions $\mathcal{P}(s)$ show an approximate scaling regime for small and intermediate relative cluster sizes $s \leq p$, cf. the vertical dashed lines in Fig.~\ref{figure8}(a), (c) and (e), followed by a cut-off due to finite system size. For the largest array size $N=200$ in Fig.~\ref{figure8}(e), we have included as a guideline the power-law scaling $\beta=3/2$ (black dash-dotted line) classically obtained for the distribution of neuronal avalanches. 

The distributions $\mathcal{P}(\tau)$ of intervals between the successive avalanche events, also called the \textit{laminar times} \cite{boffetta1999,bartolozzi2005}, indicate two different regimes that guide the avalanche recurrence processes, cf. Fig.~\ref{figure8}(b), (d) and (f). In particular, very short laminar times describe the intrinsic dynamics of cascading avalanches, i.e., correspond to cascades' intra-event intervals between the successive bursts. For intermediate $\tau$, one observes the peak that indicates the presence of a characteristic timescale in the avalanche recurrence process rather than the scale invariant behavior. Such traces of pseudo-regularity in avalanche recurrent times reflect an occasional degradation of the trapping mechanism associated with the maximal canard, which allows the system to intermittently evolve in the vicinity (not on) of the synchronization manifold, having the units generate spikes mutually shifted in phase. 

The results in this section suggest that our system in the vicinity of the transition between the regimes $R_2$ and $T_1$ from Fig.~\ref{figure2} shows certain aspects of critical behavior, like the increase of correlation length compared to coupling radius (indirectly observed by the growth of maximal cluster sizes) and the enhanced variability of cluster sizes. To further this point, in the next section we investigate the system's response to perturbations, demonstrating evidence of \emph{critical slowing down} and a \emph{decreased resilience} of the system's dynamics in the vicinity of this transition.

\subsection{Indicators of criticality} \label{indicators}

Approaching the critical transition, complex systems tend to show progressively less resilience to perturbations, taking increasingly longer times to recover \cite{Scheffer2012}. Such slower recovery rates are classically described as a herald of a critical slowing down phenomenon \cite{Meisel2015,Cocchi2017,Wilkat2019,Maturana2020}. The latter also influences the relaxation processes and hence the statistics of fluctuations underlying the spontaneous activity of systems near criticality. Qualitatively, this increases their short-term memory and variability, and is reflected in enhanced autocorrelation and variance of systems’ observables. In terms of induced activity, systems at criticality are known to maximize their dynamic range \cite{Kinouchi2006,Shew2009,Larremore2011}.

In the following, our goal is to demonstrate that at the onset of the $T_1$ region, or rather for $\sigma$ values close to the transition between regions $R_2$ and $T_1$ from Fig.~\ref{figure2}, an array of FHN units exhibits two signature effects of criticality, namely increased recovery times to small perturbations and a reduced resilience. To do so, we introduce two types of stimulation protocols: one, called an \textit{LC-shift}, where a small fraction $M$ of units is triggered to spike, i.e. their orbits are kicked toward the orbit of a relaxation oscillation limit cycle; and the other, called \textit{FP-shift}, where the same fraction of units is injected into the vicinity of the unstable fixed point $(u^*,v^*)$. The described perturbations are applied at time $t=T_p$, after which the array spontaneously evolves until the moment $t=T$. To quantify the effect of perturbations, we compare the orbit of the system after introducing the stimulus to that of the unperturbed system and numerically determine the deviations $\zeta(t)$. As a measure of the impact of the stimulus, we take the variance $\text{Var}(\zeta(t))$ of the deviations calculated over the interval $T-T_p$.

\begin{figure}[!htp]
	\begin{center}
		\includegraphics[scale=0.57]{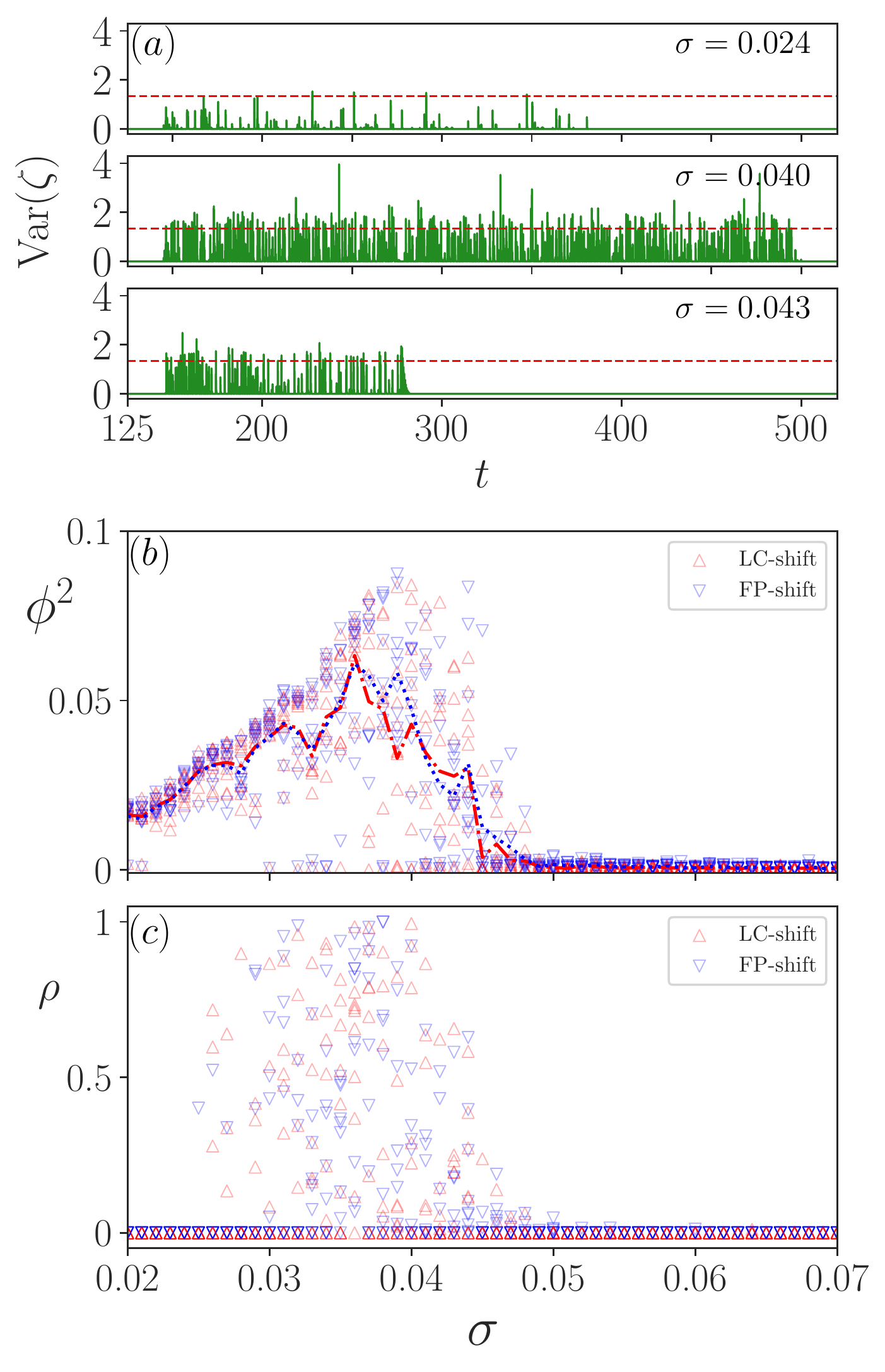}
		\caption{Indicators of criticality at the transition between regions $R_2$ and $T_1$ from Fig.~\ref{figure2}. (a) Time traces of variance $\text{Var}(\zeta)$ after an \textit{FP-shift} introduced to a fraction of $M=0.05$ units at $T_p=150$ for $\sigma=0.024$ (top panel), $\sigma=0.04$ (middle panel) and $\sigma=0.043$ (bottom panel); red dashed line: level of initial \textit{FP-shift}. (b) Cumulative variance $\phi^2$ and (c) normalized recovery time $\rho$ as a function of $\sigma$. Each symbol stands for a different set of initial conditions, and the color code refers to \textit{LC-shift}(red) and \textit{FP-shift} (blue) stimulation protocols. The dash-dotted and dotted curves in panel (b) indicate the values of $\phi^2$ averaged over an ensemble of initial conditions for \textit{LC-shift}(red) and \textit{FP-shift}(blue), respectively. System parameters: $N=50$, $p=0.2$.} \label{figure9}
	\end{center}
\end{figure}

In Fig.~\ref{figure9}(a) are shown the time series of variances $\text{Var}(\zeta(t))$ for three different values of $\sigma$ following an \textit{FP-shift} at $T_p=150$. The horizontal red dashed lines indicate the levels of the corresponding initial \textit{FP-shifts}. We first point out that the post-stimulus amplitude variance (shown green) is much higher than the initial amplitude of the \textit{FP-shift} for $\sigma=0.04$ (middle panel), whereas it is lower for $\sigma=0.024$ (top panel) and $\sigma=0.043$ (bottom panel). This reflects the array's reduced resilience, i.e. the decreased recovery capability for $\sigma=0.04$, and also shows that the perturbations from external stimuli are amplified for this value of $\sigma$. Moreover, one observes that the post-stimulus interval of nonzero variance is much longer for $\sigma=0.04$ than for the other two $\sigma$ values. This evinces that the array's recovery times $T_R$ from a perturbation (see the blue dashed line with arrows) are much slower for $\sigma=0.04$. Note that the values at the top and bottom panels are selected from regions $R_2$ and $T_1$ from Fig.~\ref{figure2}, while the longest recovery time and the largest variance amplitude are found approximately at the transition boundary between these regions. In other words, in the vicinity of the latter transition, the system shows two prominent features of criticality, having the recovery time and signal variance following a perturbation substantially different compared to the system's behavior below and above the transition.

To better characterize the described behavior, let us investigate the array's recovery times and variances over the continuous interval of $\sigma$ spanning between the regions $R_2$ and $T_1$. Our aim is to show that the variability
of the system's response to perturbations is indeed the largest in the vicinity of the transition between these two regions. Hence, for each considered value of $\sigma$, we perform simulations of the array dynamics for 10 different initial conditions and implement either the \textit{FP-shift} or the \textit{LC-shift} stimulation protocol. Then, we numerically estimate the cumulative variance per unit time $\phi^2$ for each set of initial conditions:
\begin{equation}\label{eqn:phi}
	\phi^{2} = \frac{1}{T-T_p}\,\int_{T_{p}}^{T} \text{Var}({\zeta}) dt.
\end{equation}

The dependence of the quantity $\phi^2$ on $\sigma$ is illustrated in Fig.~\ref{figure9}(b). Note that for a given value of $\sigma$, each symbol describes the system’s response for a different set of initial conditions, whereas the responses to different types of stimulation protocols are indicated by red (\textit{LC-shift}) and blue (\textit{FP-shift}). The two dotted lines indicate the system's responses averaged over the ensemble of different initial conditions for the two types of stimulus. One finds that such averaged $\phi^2$ quantities show peaks around the coupling strength $\sigma \approx 0.04$, indicating that the system is most sensitive to perturbations near the transition between the regions $R_2$ and $T_1$. Nonetheless, for the same interval of $\sigma$, we examine the array's recovery times after implementing both types of stimulation protocols. In particular, we collect the recovery times $T_R$ (indicated in Fig.~\ref{figure9}(a)) for 10 different sets of initial conditions. To make the observed values of $T_R$ comparable, we normalize them by the total observation time after the stimulus $T-T_p$, thus obtaining the normalized recovery time $\rho=T_R/(T-T_p)$. Figure \ref{figure9}(c) shows the observed values of $\rho$ as a function of $\sigma$. One readily notes that indeed the larger values of $\rho$ occur near the transition between the regions $R_2$ and $T_1$.

\section{Discussion}\label{sec:discussion}

We have introduced a simple model of an array of diffusively coupled neural oscillators whose local dynamics are poised in the vicinity of a canard transition. This facilitates a coexistence between completely synchronous oscillations and avalanche-like patterns of pseudo-synchronous bursting activity. The onset of avalanches is shown to be associated with an inhibitory effect of interactions. This effect is manifested at a range of small coupling strengths, where interactions quench local relaxation oscillations due to an interplay with a maximal canard, a structure that stems from local multiple timescale dynamics. The observed long-term trapping of orbits in the vicinity of an unstable fixed point derives from a combination of a recently introduced concept of phase-sensitive excitability of a periodic orbit \cite{FOW18} and the trapping mechanism from \cite{Medeiros2018,Medeiros2019,Medeiros2021}. Essentially, each unit, as an oscillating system driven by a fluctuating local mean-field, provides a non-uniform response to perturbations along the orbit of a limit cycle, which leads to persistent strong deviations from the unperturbed orbit. Compared to 
\cite{Medeiros2018,Medeiros2019,Medeiros2021}, the trapping phenomenon is here extended to a confinement of orbits to a region of maximal canard instead of the original confinement by a chaotic saddle. In terms of concept, one should note that distinct from the classical notion of excitability, the phase-sensitive excitability is not immediately related to the system being close to a bifurcation between stable stationary and oscillatory states, but is instead connected to a canard transition between subthreshold and relaxation oscillations. In a broader context, the important role of canard transition in pattern formation has already been shown in the cases of alternating (leap-frog) dynamics in small motifs of units \cite{EFW19} or the different types of coherence-incoherence patterns (solitary states and patched patterns) in non-locally coupled arrays with repulsive and attractive interactions \cite{Franovic2022a,Franovic2022b}, involving either coupled excitable units or self-oscillating units close to the bifurcation toward the excitable state. Complementing this, here we have shown the impact of canard transition on the self-organization and intrinsic structure of avalanche patterns. 

We have further demonstrated that avalanches can emerge at the transition between two collective regimes featuring lower and higher spiking activity rates. The avalanches have been shown to satisfy power-law behaviors regarding avalanche cluster sizes and laminar times. Moreover, the system generating avalanches has been found to bear classical indicators of criticality under external perturbations, including a reduced resilience and critical slowing down. So far, neuronal avalanches have primarily been suggested to arise in the vicinity of two very different types of continuous transitions, namely the transition between absorbing and active phases or at the onset of synchronization. Also, implementing various adaptation rules, such as synaptic plasticity or excitability adaptation, it has been indicated that models of neuronal networks may self-organize to a critical state facilitating avalanches, which has linked the onset of avalanches to self-organized criticality \cite{Levina2007,Millman2010,Rybarsch2014,Landmann2021}. At the other hand, it has been found that avalanches may emerge from critical dynamics in balanced excitatory-inhibitory networks, where they can be combined with different types of collective oscillation rhythms \cite{Poil2012,Liang2020}. The latter can involve two types of scenarios: one with collective rhythms and avalanches coexisting (either independently or with rhythms modifying the features of avalanches), and the other having the rhythms embedded in avalanche activity \cite{Gireesh2008,Lombardi2014}. Finally, it has been reported that scale-invariant avalanches may also emerge without the neural network operating at criticality, but just due to a balanced input or its interaction with noise \cite{Lombardi2012,Poil2012,Gautam2015,Liang2020}.

In light of the above studies, our findings apparently point to a possibility of an independent coexistence between a synchronous oscillation rhythm and transiently synchronous avalanche activity, whereby the mechanism facilitating such coexistence requires two ingredients: the non-local diffusive interactions and local dynamics in the vicinity of a canard transition between subthreshold and relaxation oscillations. In terms of the states involved, the character of the critical transition supporting avalanches is most similar to the one in \cite{Scarpetta2018}, in the sense that it also mediates between the states with lower- and higher spiking rates. Nevertheless, in contrast to our study, the model in \cite{Scarpetta2018} has a more complex structure combining stochastic local dynamics with a quenched disorder in network topology, and criticality occurs in the vicinity of a spinodal line of a discontinuous transition. For future research, it would be important to gain insight into the switching dynamics between the coexisting regimes in our model, both under the impact of noise and when applying different types of external stimulation.

\begin{acknowledgments}
\noindent M.C. thanks Javiera Contreras for the support and help with the design of the figures. \newline
E.S.M. acknowledges the support by the Deutsche Forschungsgemeinschaft (DFG) via the project number 454054251. This work was supported by the Deutsche Forschungsgemeinschaft (DFG, German Research Foundation) - Projektnummer - 163436311 - SFB 910.\newline
A.Z. acknowledges support by the Deutsche Forschungsgemeinschaft (DFG, German Research Foundation) - Projektnummer - 163436311 - SFB 910.\newline
P.H. acknowledges support by the Deutsche Forschungsgemeinschaft (DFG, German Research Foundation) – Project-ID 434434223 – SFB 1461.\newline
I.F. acknowledges the funding from the Institute of Physics Belgrade through grant by the Ministry of Science, Technological Development and Innovation of the Republic of Serbia and the partial support by the ANSO - Alliance of International Science Organizations Collaborative Research Projects and Training Projects, grant number ANSO-CR-PP-2022-05.
\end{acknowledgments}

\section*{Data Availability}
The data that support the findings of this study are available from the corresponding author upon reasonable request.

\bibliography{fhn}

\end{document}